\documentclass{ws-procs961x669}  
\usepackage{mathtools}          
\begin{document}
\title{Nonperturbative Casimir effects:
Vacuum structure, Confinement, and Chiral Symmetry Breaking
}

\author{A.V.Molochkov}

\address{Pacific Quantum Center, Far Eastern Federal University, Street,\\
Vladivostok, 690001, Russian Federation\\
E-mail: molochkov.alexander@gmail.com}

\begin{abstract}
The review of vacuum and matter restructuring in space-time with boundaries is presented. We consider phase properties of confining gauge theories and strongly interacting fermion systems. In particular, the chiral and deconfinement phase transitions properties in the presence of Casimir plates. We also discuss mass scale shifts in such systems and their possible dynamical and geometrical nature.  
\end{abstract}


\bodymatter

\section{Introduction}

One of the fundamental problems of quantum field theory is the consistency of non-trivial geometry and quantization. One of the brightest examples in this area is quantum fields in space with boundaries. In the case of a pure vacuum, certain boundaries give rise to Casimir forces~\cite{Casimir:1948dh}. Recent theoretical works have shown the possibility of vacuum restructuring effects occurring in the systems with boundaries, leading to a change in the gauge theory vacuum's phase properties and critical behaviour. In particular, it can lead to the low-temperature deconfinement phase transition~\cite{Chernodub:2017mhi, Chernodub:2017gwe, Chernodub:2018pmt}. One can also assume that the effects of space-time boundaries should manifest themselves in systems with matter fields, particularly in fermionic systems~\cite{Flachi:2013bc, Flachi:2012pf, Chernodub:2016kxh, Tiburzi:2013vza}. The fundamental question is - what is the nature of these effects? Is it a consequence of the dynamics of the restructured vacuum, or is it a consequence of the interaction with the Casimir boundaries, or is it a result of a finite space-time volume? 

Since its discovery, the Casimir effect has been intensively studied experimentally and theoretically~\cite{ref:Bogdag,ref:Milton}. The experimental studies~\cite{Lamoreaux:1996wh, Mohideen:1998iz} confirmed this effect in plate-sphere geometries in agreement with theory. 

In his original work, Hendrik Casimir assumed that the vacuum gauge field's state spectrum of a system with boundaries is limited in the infrared region due to finite size, which leads to the observable difference in energy densities of the unbound vacuum and the vacuum with boundaries. As a result, the attracting forces between the boundaries arise. However, several theoretical results indicate a non-trivial change in the vacuum structure, which cannot be explained by an infrared restriction of the vacuum spectrum. 

In particular, a spherical geometry leads to the repulsive Casimir force, which is acting outwards~\cite{Boyer}
\begin{equation}
	<E_{Sphere}>= +\frac{0.0461765}{R}
\end{equation}

Another interesting theoretical result was that Scharnhorst found that in the Casimir vacuum in the low-frequency region $\omega \ll m$, where $m$ is the electron mass, light propagation modes have phase velocity exceeding $c$~\cite{Scharnhorst:1990sr}. 

In the paper~\cite{Scharnhorst:1993}, Scharnhorst and Barton assumed that the Casimir vacuum behaves like a passive medium ($Im\,(n_{\perp}(\omega)) \ge 0$). This assumption led them to the conclusion that the front velocity of light $c/n_{\perp}(\infty)$ in the Casimir vacuum is greater than the speed of the light front $c$ in the unbound vacuum. At the same time, the author emphasises that this conclusion does not contain any severe conceptual dangers and, in particular, does not contradict the special theory of relativity.  

These examples show two different sources of the Casimir vacuum change. The first example shows that the geometry of the Casimir boundaries significantly affects the properties of the vacuum. The second shows fundamental physical scales change due to the non-perturbative Casimir dynamics and the final size of the system. 

In the first part of the paper, we will briefly review the non-perturbative properties and phase structure of field theories with plain Casimir boundaries (for a detailed review see the paper~\cite{Chernodub:2019nct}. In the second part, we will discuss the effects of the four-dimensional geometry of bounded space-time to understand its relations with mass-scale shifts in finite volume systems.

\section{Phase structure of field theories in space with boundaries}
This section considers three examples of the scale shift in systems with Casimir boundaries - strongly interacting fermionic system, compact electrodynamics, and SU(2)theory. 

{\bf Strongly interacting fermionic system.}
The critical phenomenon of strong interactions is the spontaneous breaking of chiral symmetry, which occurs in the fermionic QCD sector. Chiral symmetry breaking in systems with Casimir boundaries was studied in the model of interacting fermions with the Lagrangian~\cite{Flachi:2013bc,Flachi:2012pf}: 
\begin{equation}
	{\cal L}= i\bar\Psi\not\partial\Psi+\frac{g}{2}(\bar\Psi\Psi)^2
\label{eq:L:GN}\end{equation}
This theory is invariant under discrete $Z_2$ chiral transformations $\Psi\to \gamma_5\Psi$ of a fermionic field with N-flavors. 
\begin{figure}[!thb]
\vskip 1mm
\begin{center}
\includegraphics[scale=0.11,clip=true]{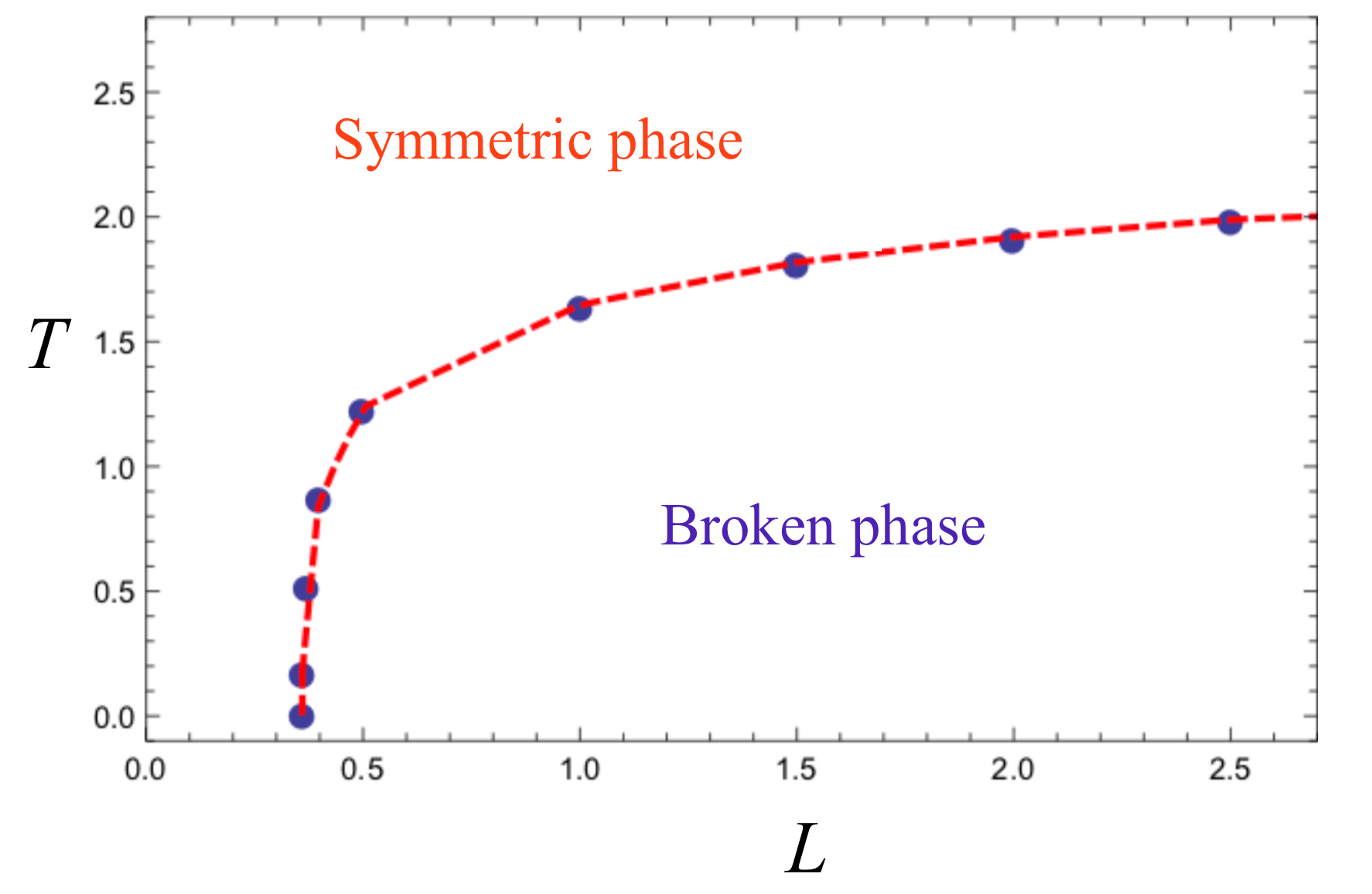}
\end{center}
\caption{The phase diagram of the (3+1) dimensional model of interacting fermions~(\ref{eq:L:GN}) for the inter-plate distance $L$ and temperature $T$, from Ref.~\cite{Flachi:2013bc}. The dimensional quantities are expressed in units of the coupling $g$.}
\label{fig:fermionic}
\end{figure}

In the unbound space, the vacuum of this model forms a dynamic chiral condensate $\langle\bar\Psi\Psi\rangle$ that breaks the chiral symmetry. Symmetry is restored at high temperatures through a second-order phase transition. The critical temperature decreases in the presence of Casimir plates, and the phase transition becomes first order.
The chiral symmetry is restored at a sufficiently small distance between the plates, even at zero temperature. Figure~\ref{fig:fermionic} shows the phase diagram of this model. The restoration of chiral symmetry due to the Casimir geometry agrees well with the observation that boundary effects restore chiral symmetry in the broken phase~\cite{Tiburzi:2013vza, Chernodub:2016kxh}.

Thus, the presence of the Casimir boundaries strongly affects field theory's critical behaviour and symmetry properties.  

Let us consider the non-perturbative analysis of the symmetry breaking in confining theories - compact electrodynamics and SU(2) gauge theory. 

{\bf Compact eletrodynamics}
Compact electrodynamics is another toy model with interesting nonperturbative properties similar to QCD. It has linear confinement of electric charges and the presence of non-trivial topology in physically significant cases of two and three spatial dimensions.

Below, we will briefly consider the compact QED's vacuum structure transformation due to the Casimir boundaries' presence. The presented analysis was performed within the first-principles simulations of lattice field theory.
 The technical details and definitions can be found in the papers~\cite{Chernodub:2017mhi, Chernodub:2017gwe, Chernodub:2018pmt}.  

The important feature of the compact QED is the presence of monopole singularities. 
In two spatial dimensions, the monopole is an instanton-like topological object that arises due to the compactness of the gauge group.

The presence of monopoles generates the mass gap
\begin{equation}
	m=\frac{2\pi\sqrt{\rho}}{g}
\label{mass_gap}\end{equation}
and a finite-temperature phase transition at a certain critical temperature $T=T_c$. Here $\rho$ is monopole density, and $g$ is the lattice coupling.

In two spatial dimensions, the standard Casimir boundary conditions are formulated for one-dimensional objects (“wires”). A static and infinitely thin wire, made of a perfect metal, forces the tangential component of the electric field ${\bf E}$ to vanish at every point $x$ of the wire, $E_{\parallel}(x) = 0$. The wire does not affect the pseudoscalar magnetic field $B$. 
\begin{figure}[!thb]
\begin{center}
\begin{tabular}{cc} 
\includegraphics[scale=0.4,clip=true]{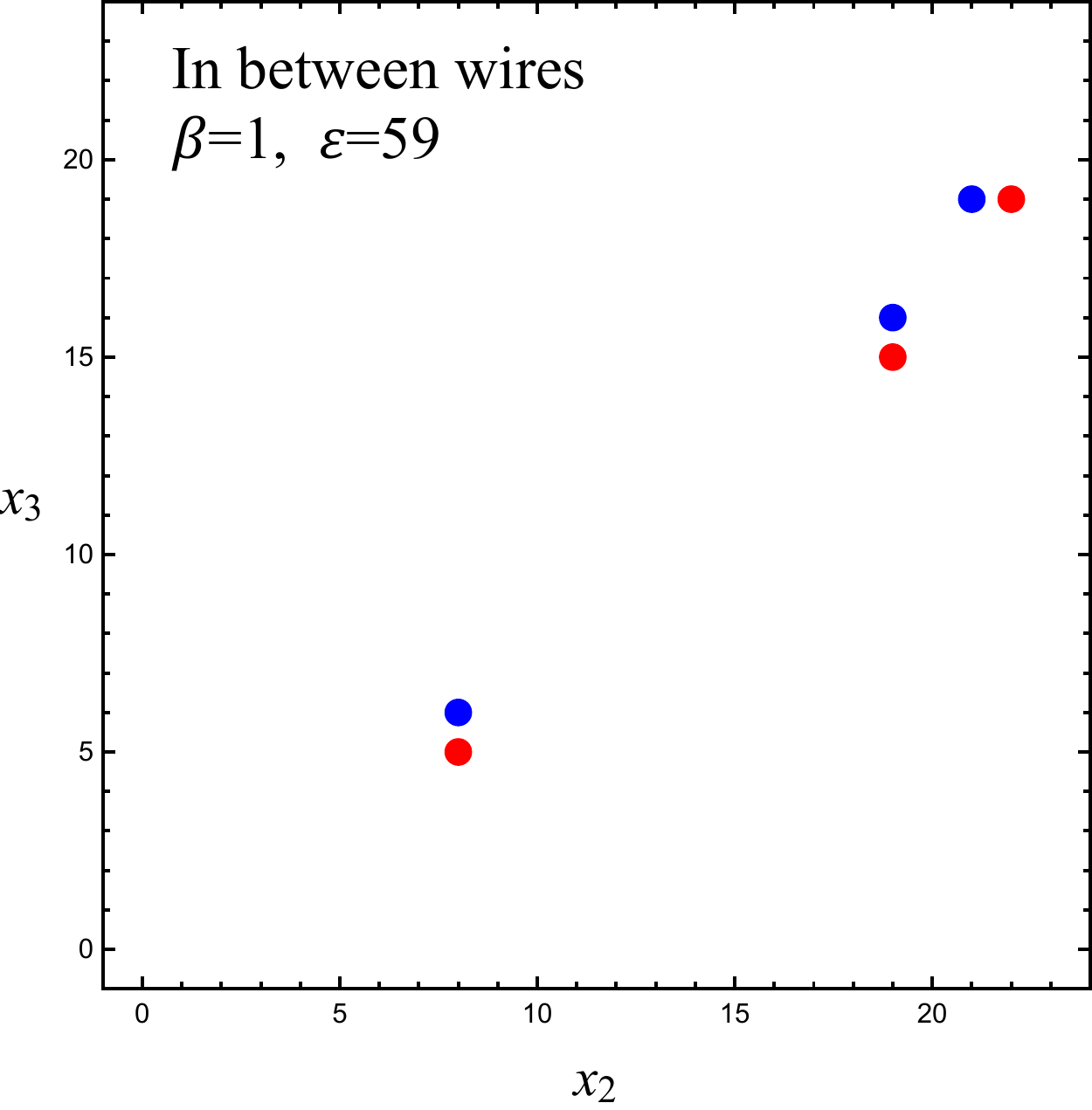} & 
\includegraphics[scale=0.4,clip=true]{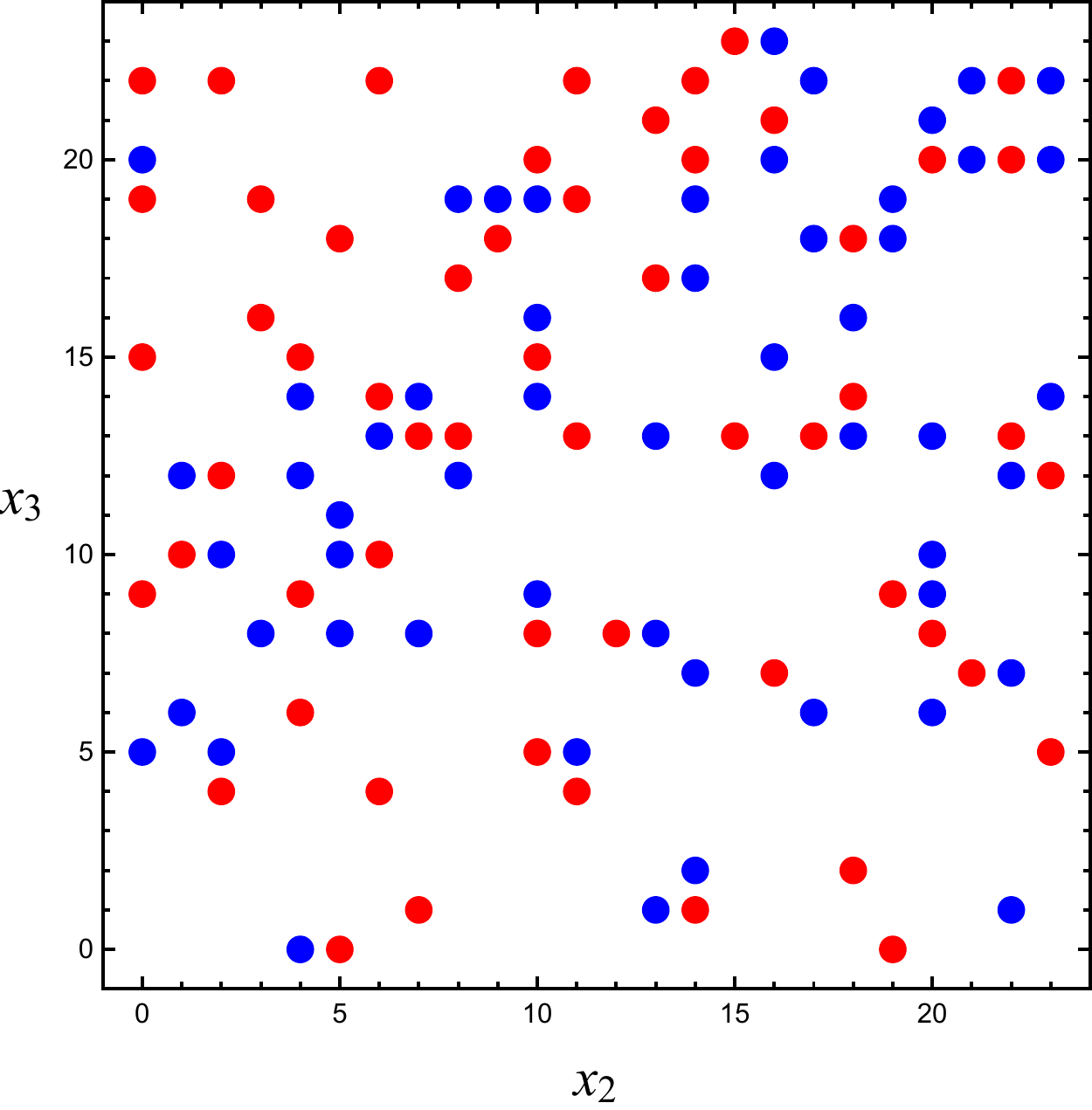} \\
 (a) & (b)
\end{tabular}
\end{center}
\caption{
Examples of typical configurations of monopoles (blue) and antimonopoles (red) in (a) a slice in between closely spaced plates and (b) in a space outside the plates
(from \cite{Chernodub:2017mhi}).}
\label{fig:geometry:plane}
\end{figure}
The Casimir energy density corresponds to the component of the canonical energy-momentum tensor $T^{00} = (E^2 + B^2)/(2g^2)$. Numerical calculations show that the presence of Abelian monopoles has a nonperturbative effect on the Casimir effect~\cite{Chernodub:2017mhi}. The mass gap~(\ref{mass_gap}) screens the Casimir energy at large distances between the wires. At small distances, it is the wires that act on the monopoles. As the wires approach, the relatively dense monopole gas between them is continuously transformed into a dilute gas of monopole-antimonopoly pairs, as shown in Fig.~\ref{fig:geometry:plane}(a) and (b)~\cite{Chernodub:2017gwe}. The geometry-induced binding transition is similar to the Berezinsky--Kosterlitz--Thouless (BKT)-type infinite-order phase transition~\cite{ref:BKT} that occurs in the same model at a finite temperature~\cite{ref:binding:cU1}.
 
\begin{figure}[!thb]
\begin{center}
\begin{tabular}{cc}
\includegraphics[scale=0.45,clip=true]{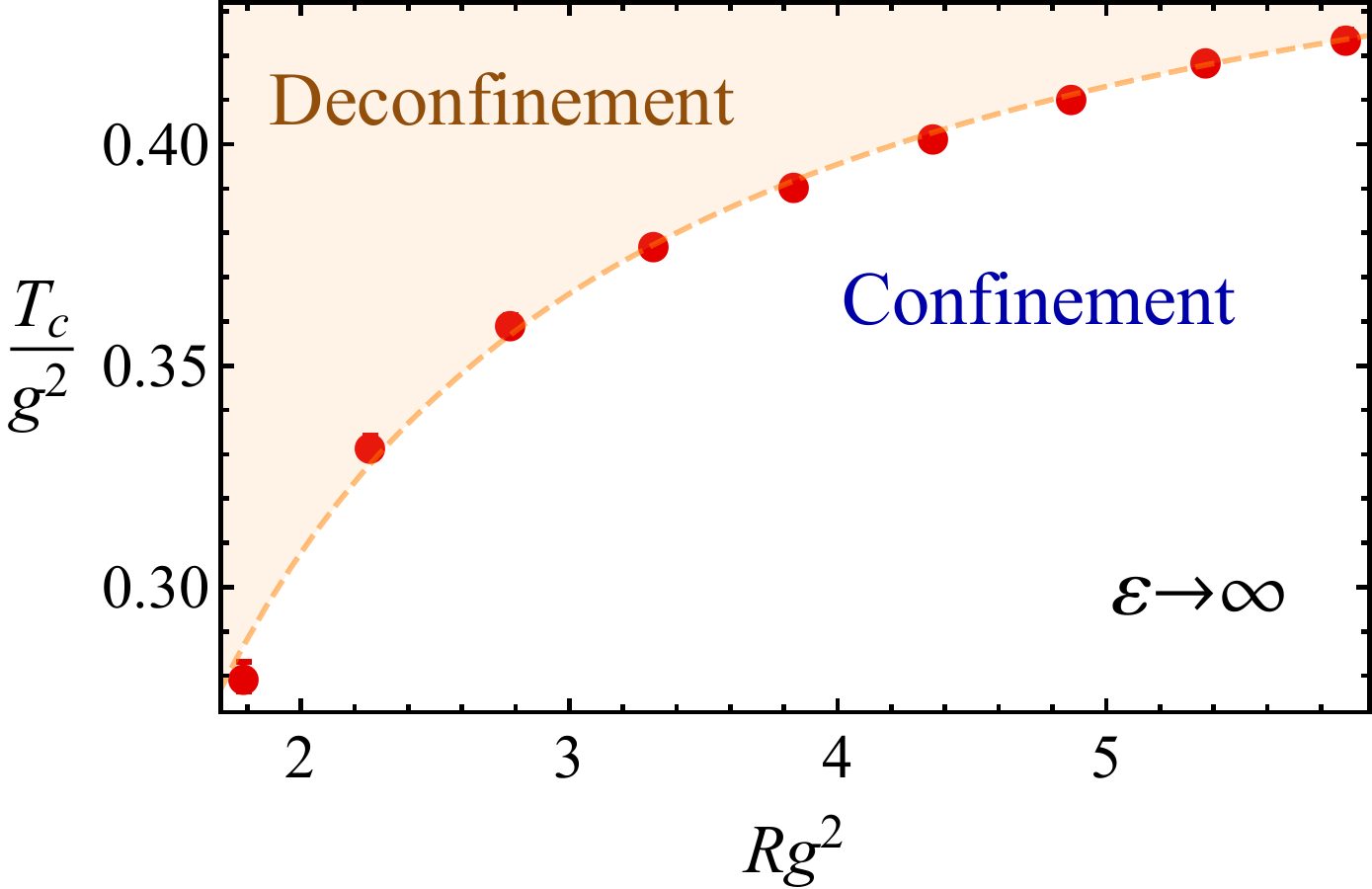} & 
\hskip 5mm
\includegraphics[scale=0.08,clip=true]{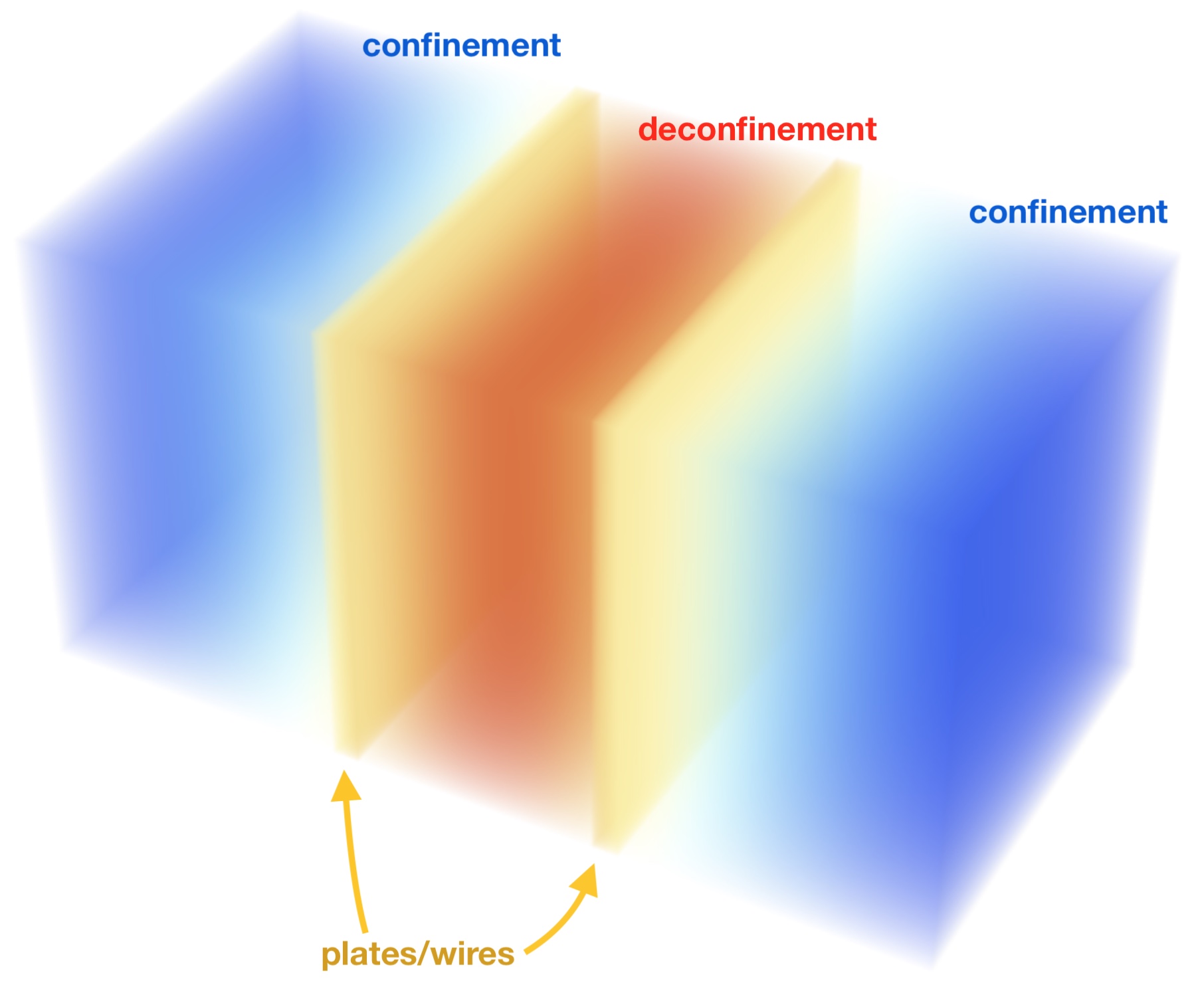} \\
(a) & (b)
\end{tabular}
\end{center}
\caption{(a) The phase in-between the plates: the critical temperature $T_c$ of the deconfinement transition as the function of the inter-plate distance~$R$ in units of the electric charge $g$ in the ideal-metal limit ($\varepsilon \to \infty$). (b) An illustration of the deconfinement in the space between the plates (from Ref.~\cite{Chernodub:2017gwe}).}
\label{fig:deconfinement}
\end{figure}

 The BKT transition is associated with a loss of the confinement property between the metallic plates because the weak fields of the magnetic dipoles cannot lead to a disorder of the Polyakov-line deconfinement order parameter. This conclusion agrees well with expectation with a direct evaluation of the Polyakov line in between the plates~\cite{Chernodub:2017gwe}. Figure~\ref{fig:deconfinement}(a) shows the phase structure of the vacuum of compact electrodynamics in the space between long parallel Casimir wires at finite temperature $T$. The deconfinement temperature $T_c$ is a monotonically rising function of the interwire distance~$R$. Formally, the charge confinement disappears completely when the separation between the plates becomes smaller than the critical distance $R = R_c$ determined by the condition $T_c(R_c) = 0$. According to the numerical estimates~\cite{Chernodub:2017gwe}, $R_c = 0.72(1)/g^2$\,.
 
{\bf SU(2) theory.} 
Let us consider the Casimir effect for a non-Abelian gauge theory which possesses an inherently nonperturbative vacuum structure. It is instructive to consider a zero-temperature Yang-Mills theory in (2+1) spacetime dimensions. 
The model in (2+1) dimensions exhibits mass gap generation and colour confinement similarly to its 3+1 dimensional counterpart. 

\begin{figure}[!thb]
\begin{center}
\begin{tabular}{cc}
\includegraphics[scale=0.05,clip=true]{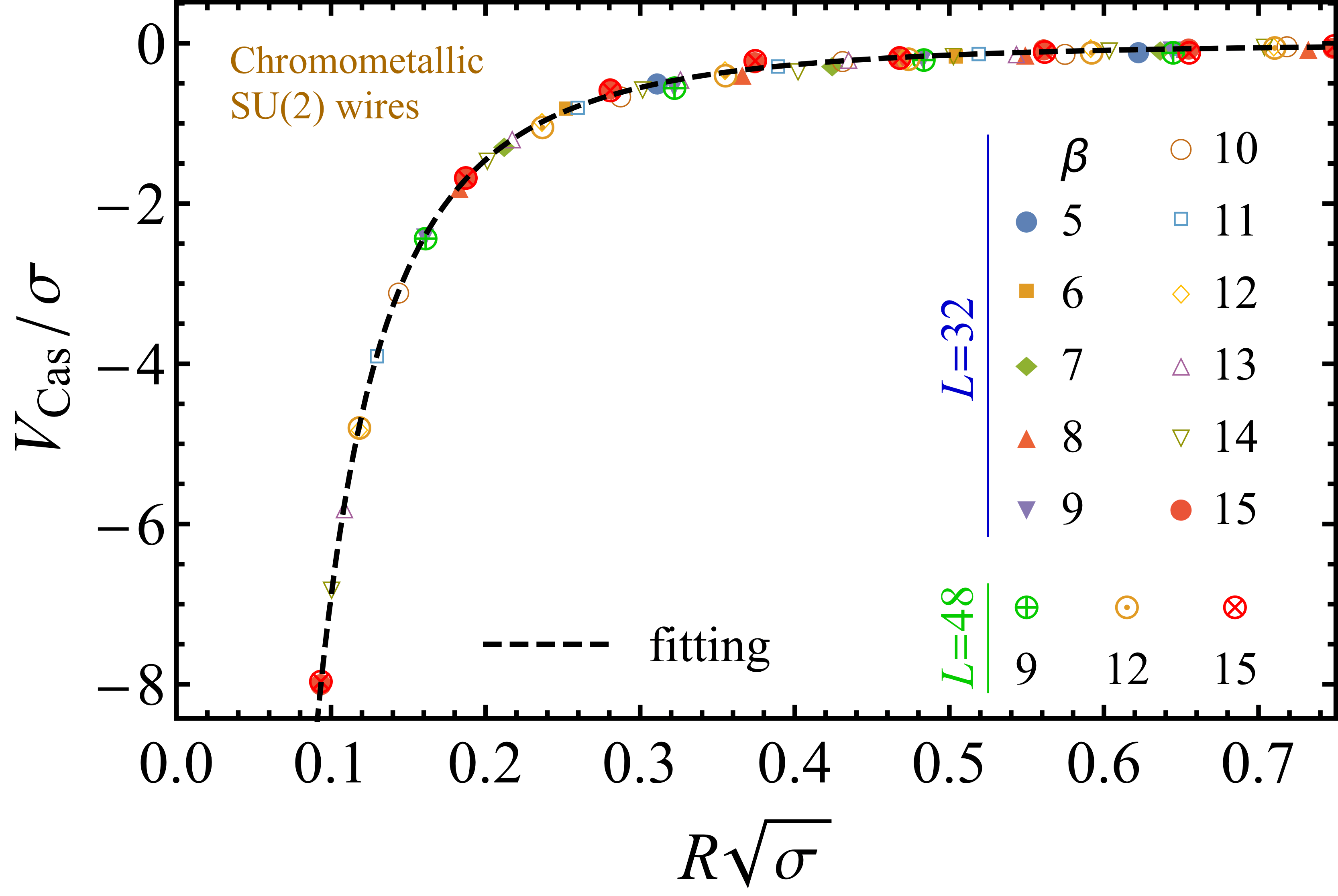} & 
\includegraphics[scale=0.57,clip=true]{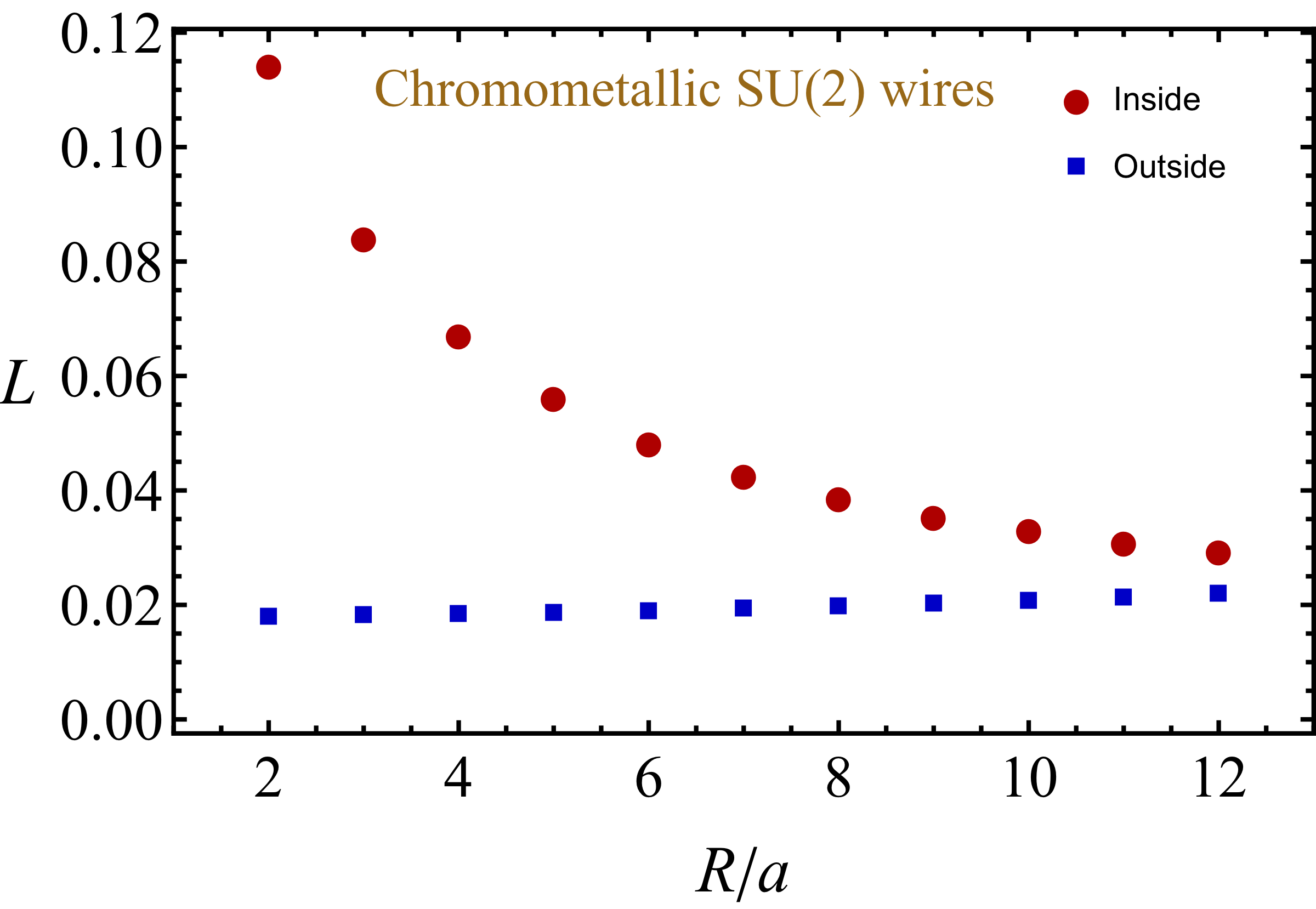} \\
(a) & (b)
\end{tabular}
\end{center}
\caption{(a) The Casimir potential $V_{Cas}$ for a chromo-metallic wire as the function of the distance $R$ between the wires (in units of the string tension~$\sigma$) at various ultraviolet lattice cutoffs controlled by the lattice spacing~$\beta$. The line is the best fit~(\ref{eq:V:fit}). (b) A typical expectation value of the absolute value of the mean Polyakov line in the spaces in between and outside the wires vs the interwire separation~$R$~\cite{Chernodub:2018pmt}.}
\label{fig:V}
\end{figure}

The Casimir energy of gluon fluctuations per unit length of the wire is shown in Fig.~\ref{fig:V}(a). The lattice results -- which exhibit excellent scaling with respect to a variation of the lattice cutoff -- can be described very well by the following function:
\begin{equation}
	V_{Cas}(R) = 3 \frac{\zeta(3)}{16 \pi} \frac{1}{R^2}\frac{1}{(\sqrt{\sigma} R)^{\nu}} e^{- M_{Cas} R},
\label{eq:V:fit}
\end{equation}
where the anomalous power $\nu$ (which controls the short-distance behaviour) and the ``Casimir mass'' $M_{Cas}$ (which is responsible for the screening at large inter-wire separations). The values $\nu = 0$ and $M_{Cas} = 0$ correspond to the Casimir energy of three non-interacting vector particles. In the SU(2) Yang-Mills theory, one gets:
\begin{equation}
	M_{Cas} = 1.38(3) \sqrt{\sigma}\,, 
\qquad
\nu_\infty = 0.05(2).
\label{eq:M:infty}
\end{equation}
Surprisingly, the Casimir mass $M_{Cas}$ turns out to be substantially smaller than the mass of the lowest colorless excitation, the $0^{++}$ glueball, $M_{0^{++}} \approx 4.7 \sqrt{\sigma}$  (the latter quantity has been calculated numerically in Ref.~\cite{Teper:1998te}). In Ref.~\cite{Karabali:2018ael} it was shown that the (2+1) Casimir mass might be related to the magnetic gluon mass of the Yang-Mills theory in (3+1) dimensions.

In Fig.~\ref{fig:V}(b), we show the expectation value of the order parameter of the deconfining transition, the Polyakov loop $L$, in the space inside and outside the wires. Similarly to the compact Abelian case, the gluons in between the wires experience a smooth deconfining transition as the wires approach each other, thus confirming the qualitative picture shown in Fig.~\ref{fig:deconfinement}(b).

\section{Effects of the four-dimensional boundaries}

We considered systems with time-like space boundaries that lead to non-trivial properties and vacuum structure change. Another essential configuration of the boundaries is space-like temporal hyper-surfaces that lead to finite temporal size systems. By the analogy of a space box that as a whole has space-translation symmetry, we can consider a finite temporal box that as a whole has continuous time-translational symmetry. It corresponds to the space-time box, which is at the rest frame. Here we will show how the relativistic treatment of the bound state allows us to consider it a space-time box. We will also discuss the physical outcomes of the existence of the temporal boundaries. 

Let us consider a relativistic bound state of two particles.
 The bound particles are distributed in the bound state's internal space-time (space-time bag). Instead of the traditional formulation of the Casimir effect, where borders are externally set boundary conditions, the border here is of dynamical origin. The boundaries result from the bound particles' dynamics, which naturally leads to the vanishing the current's radial component at the bound state boundaries. This condition is equivalent to the MIT bag condition. 
 
The bound state of particles can be defined as the constituents scattering amplitude pole in the momentum space:
\begin{equation}
	G(p_1, p_2, p^\prime_1, p^\prime_2) = \frac{\Gamma(p_1,p_2)\bar\Gamma(p^\prime_1, p^\prime_2)}{(p_1+p_2)^2-M^2} + ....,
\label{pole}\end{equation}
where $M$ is the mass of the bound state, $p_1,\,p_2$ - 4-momenta of the constituents.

Below, we will discuss the relation between the appearance of the pole and the finite temporal size of the bound state, which in its order is related to the shifting of the particles from the mass shell.

For this purpose, let us consider the 4D kinematics of the two-particle system. To separate centre of mass constituents kinematics, instead of individual 4-momenta of the particles ($p^\mu_1, p^\mu_2$), we will use the set of 4-momenta of the centre of mass and the relative momenta of the particles ($P^\mu, k^\mu$): 
\begin{equation}
	P^\mu=p^\mu_1+p^\mu_2,\,\,\,\ 2k^\mu=p^\mu_1-p^\mu_2
\end{equation}
Let us consider, for simplicity, the centre of mass rest frame. In this case the momenta $p^\mu_1$ and $p^\mu_2$ can be written in the form: 
\begin{eqnarray}
p^0_1=\frac{M}{2}+k_0, \,\,\,\, p^0_2=\frac{M}{2}-k_0	\\\nonumber
{\mathbf p}_1=-{\mathbf p}_2={\mathbf k}
\label{kinCM}
\end{eqnarray}
The corresponding space-time variables have the following form:
\begin{eqnarray}
x^0_1=T+\frac{\tau}{2}, \,\,\,\, x^0_2=T-\frac{\tau}{2}	\\\nonumber
{\mathbf x}_1={\mathbf X}+\frac{\mathbf \chi}{2}, \,\,\,\, {\mathbf x}_2={\mathbf X}-\frac{\mathbf \chi}{2},
\label{XkinCM}
\end{eqnarray}
 where ($T$, ${\bf X}$) - center of mass space-time, ($\tau$, ${\bf \chi}$) - space-time of constituents relative positions.
 
The relative energy $2k_0\ne 0$ for the off-shell particles system only.
Indeed, the on-shell conditions in terms of the variables~(\ref{kinCM}) are the following:
\begin{eqnarray}
	\frac{M}{2}+k_0=E({\bf k})\,\,\,\,\,\,\,\,
	\frac{M}{2}-k_0=E({\bf k})\,\,\,\,\,\,\,\,
	E({\bf k})=\sqrt{{{\bf k}^2}+m^2}\label{ms_1}
\end{eqnarray}
The solution of the system gives the following mass-shell condition:
\begin{equation}
	k_0=0,\,\,\, M=2E({\bf k})\ge 2m\label{mass_shell}
\end{equation}
Thus, $k_0$ determines the degree of the nucleon's off-mass-shell shift. Below, we will show that the finite temporal size of the bound state leads to $k_0\neq 0$.

Let us consider the Green function of a bound state of two particles:
 \begin{equation}
 	 G(P) = \int\frac{d^4k}{(2\pi)^4}\frac{\bar\Gamma(P,k)\Lambda_1(P,k)\otimes\Lambda_2(P,k)\Gamma(P,k)}{\left[\left(\frac{P}{2}+k\right)^2 -m^2\right]\left[\left(\frac{P}{2}-k\right)^2-m^2 \right]},
\label{BS_GreenF} \end{equation}
where 
$$\Lambda_1(P,k)=\left(\frac{P}{2}+k\right)\gamma+m,\,\,\,\, \Lambda_2(P,k)=\left(\frac{P}{2}-k\right)\gamma++m.$$

\begin{figure}[!thb]
\begin{center}
\begin{tabular}{cc}
\includegraphics[scale=0.15,clip=true]{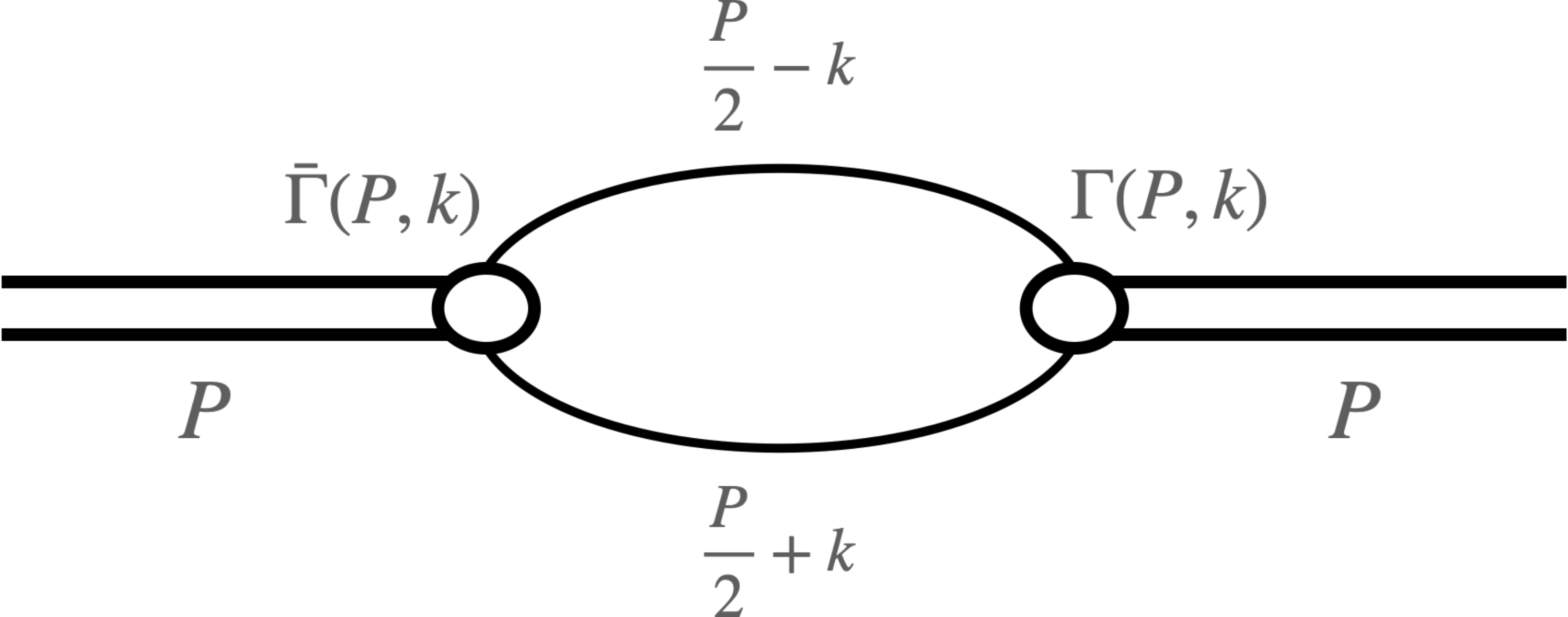}
&\includegraphics[scale=0.19,clip=true]{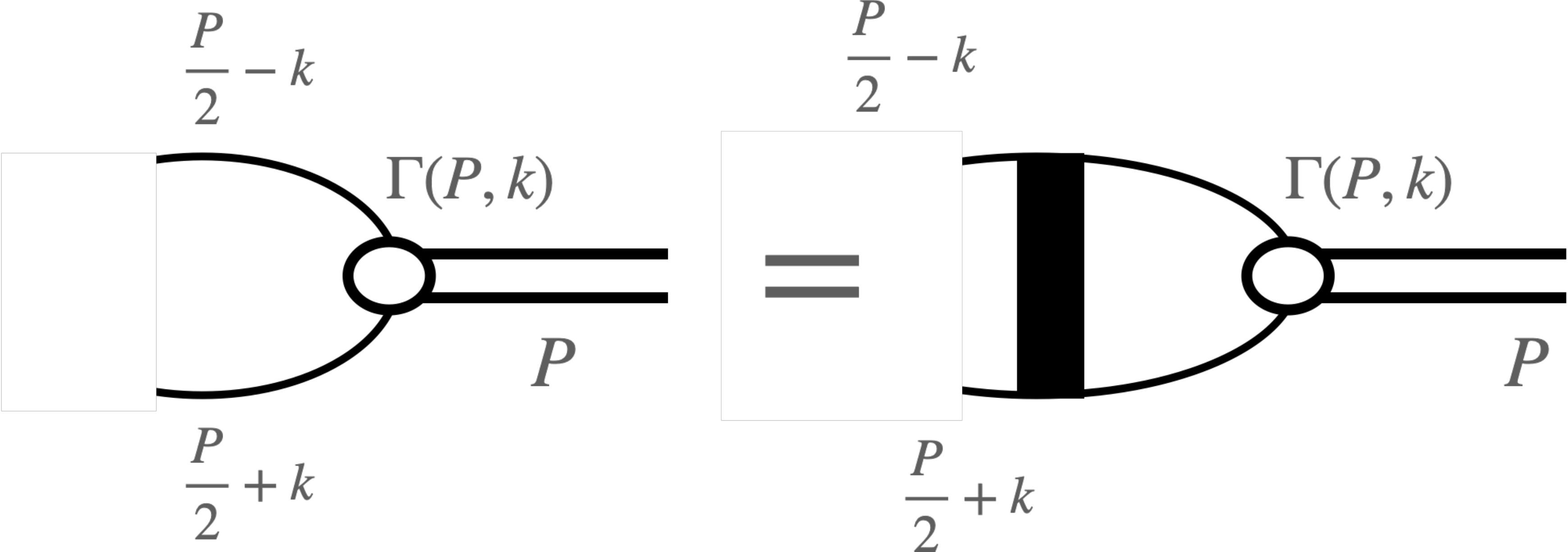}\\
(a) & (b)
\end{tabular}
\end{center}
\caption{(a) The diagrammatic illustration of the bound state propagator. \\ (b) The diagrammatic illustration of the Bethe-Salpeter equation.}
\label{BS_diagramm}
\end{figure}

Figure~\ref{BS_diagramm} illustrates this Green function by the corresponding Feinman diagram.  
It contains contributions of positive and negative energy states. Let us consider positive energy contributions. The negative energy terms can be analysed in the same way. 

The positive term contribution to the Eq.(\ref{BS_GreenF}) has the following form:
\begin{equation}
	G^{++}(P) = \int\frac{d^4k}{(2\pi)^4}\frac{\bar\Gamma^{++}(P,k)[u({\bf k})\bar u({\bf k})\otimes u(-{\bf k})\bar u(-{\bf k})]\Gamma^{++}(P,k)}{4E^2\left[\frac{M}{2}+k_0-E\right]\left[\frac{M}{2}-k_0-E\right]}
\label{BS_PP}\end{equation}  

The initial and final total 4-momenta $P, P^\prime$ are equal to each other due to momentum conservation. This term considers any intermediate interaction corrections due to the Bethe-Salpeter equation (see Fig~\ref{BS_diagramm}b. Thus, the relative 4-momenta $k$ are conserved too. 

The integrand of Eq.(\ref{BS_PP}) represents the momentum distribution of the particles within the bound state. The total momentum $P$ dependence is trivial in the case of ${\bf P}=0$.
 The corresponding internal space-time distribution can be obtained as a Fourie transformation of the integrand. 
 \begin{equation}
 	g(\tau,{\mathbf \chi}) = \int\frac{d^4k}{(2\pi)^4}\frac{-f(k) e^{ik_0\tau-i{\bf k}\cdot{\bf \chi}}}{4E^2(k_0-(E-\frac{M}{2}))(k_0+(E-\frac{M}{2}))},
\label{DistFF} \end{equation}
where
$$f(k)=\bar\Gamma^{++}(M, k)[u({\bf k})\bar u({\bf k})\otimes u(-{\bf k})\bar u(-{\bf k})]\Gamma^{++}(M,k).$$
 
Let's integrate~(\ref{DistFF}) with respect $k_0$ taking into account the poles $k_0=\pm(E-\frac{M}{2})$:

\begin{eqnarray*}
&&g(\tau,{\bf \chi})=\int\frac{d^3{\bf k}}{(2\pi)^3}f\left(E-\frac{M}{2}, {\bf k}\right)e^{-i{\bf k}\cdot{\bf\chi}}	\frac{1}{i2E^2}\left[\frac{e^{i(E-\frac{M}{2})\tau}}{2(2E-M)}-\frac{e^{-i(E-\frac{M}{2})\tau}}{2(2E-M)}\right]=\\
&&=\int\frac{d^3{\bf k}}{(2\pi)^3}f\left(E-\frac{M}{2}, {\bf k}\right)e^{-i{\bf k}\cdot{\bf \chi}}	\frac{sin((2E-M)\tau)}{2E^2(2E-M)}
\end{eqnarray*} 
 We calculate the Fourie image with respect ${\bf k}$ as a Fourie transform of the product of the functions:
 \begin{equation}
 g(\tau,{\bf \chi})=\int d^3{\bf\chi^\prime} g_1({\bf \chi}-{\bf \chi^\prime})g_2(\tau, {\bf \chi^\prime})	
 \label{Full_g}\end{equation}
 where 
 \begin{equation}
 	g_1({\bf \chi}- {\bf \chi^\prime})=\int\frac{d^3{\bf k}}{(2\pi)^3} f(2E-M, {\bf k})e^{-i{\bf k}\cdot ({\bf \chi}-{\bf \chi^\prime})} 
 \end{equation}
 and
 \begin{eqnarray}
 	g_2(\tau,{\bf \chi}^\prime)=\int\frac{d^3{\bf k}}{(2\pi)^3} \frac{sin((E-\frac{M}{2})\tau)}{2E^2(2E-M)}e^{i{\bf k}\cdot{\bf \chi}^\prime}.
 \label{g2}\end{eqnarray}
 The function $g_2$ fully desribes the $\tau$-dependence of the distribution function $g(\tau, {\bf \chi})$.
  It does not depend explicitly on the constituents' interaction and has a pure geometrical nature. According the expression~(\ref{Full_g}) it plays role of function limiting the three dimensional space distribution $g_1$. Thus, one can consider it as a kind of integral boundary condition.
  
 Calculation of the integral in~(\ref{g2}) with respect ${\bf k} $ gives the temporal distribution of the constituents within the bound state:
 \begin{equation}
 	g_2(\tau,{\bf \chi^\prime})=\int\limits_0^\infty |{\bf k}|^2d|{\bf k}| \frac{sin((E-\frac{M}{2})\tau)}{E^2(2E-M)} \frac{sin(|{\bf k}||{\bf \chi}^\prime|)}{|{\bf k}||{\bf \chi}^\prime|} 
\end{equation}

\begin{figure}[!thb]
\begin{center}
\begin{tabular}{cc}
\includegraphics[scale=0.39,clip=true]{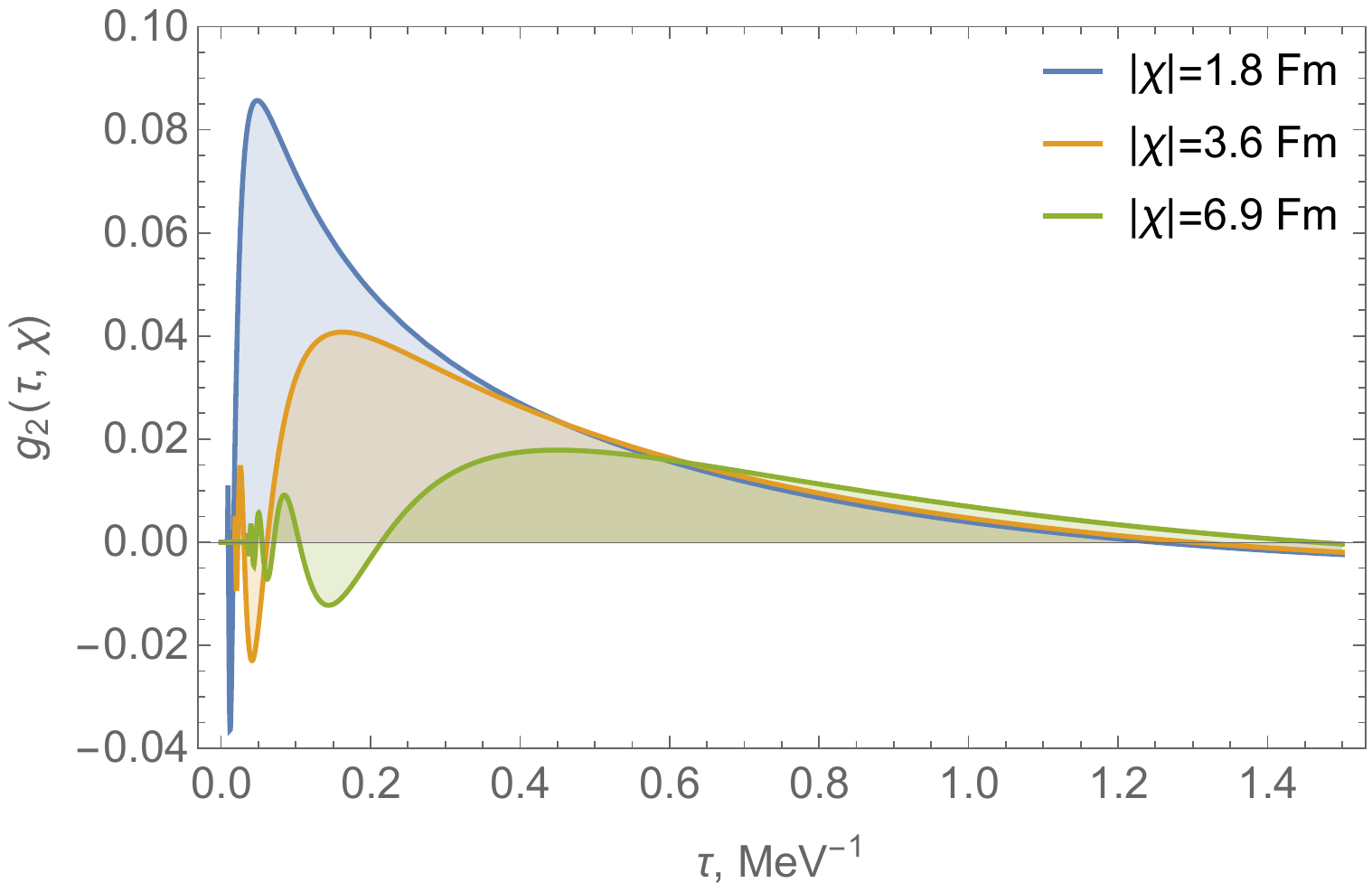}
&\includegraphics[scale=0.39,clip=true]{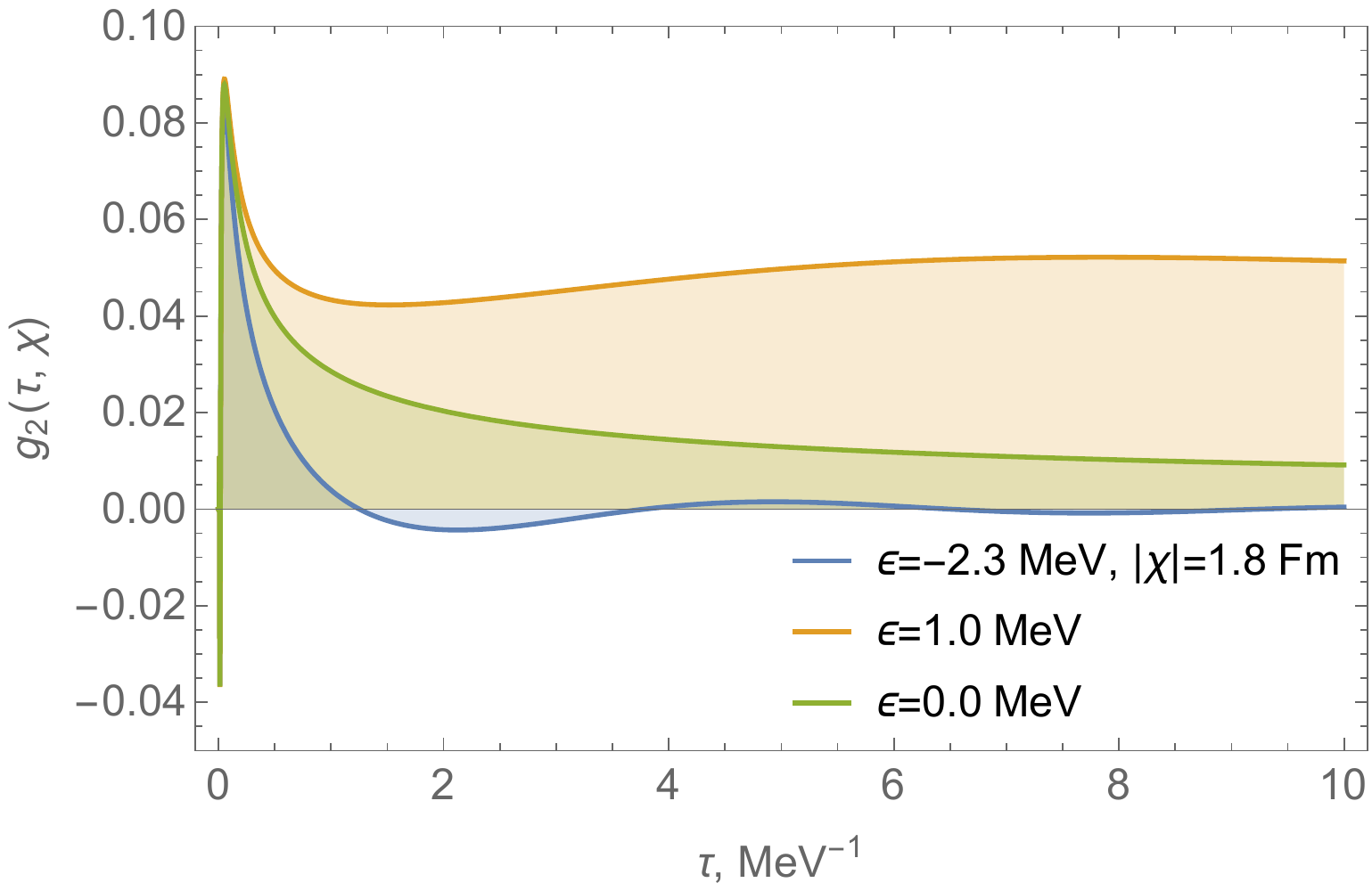}\\
(a) & (b)
\end{tabular}
\end{center}
\caption{(a) The temporal distribution $g_1(\tau,\chi)$ for the deuteron at three different values of space distance $\chi=1.8 Fm$ (blue line), $\chi=3.6 Fm$ (orange line), $\chi=6.9 Fm$ (green line). \\ (b) The temporal distribution $g_2(\tau,\chi)$ for the deuteron-like bound state with different values of the binding energy at $|\chi|=1.8Fm$.}
\label{D_large_t}
\end{figure}

To illustrate a relation between temporal structure and energy scales, we choose values of $m$ and $M$ of a physical bound state of two nucleons - deuteron. The calculation result is presented in the figure~\ref{D_large_t}a.

The Fig.~\ref{D_large_t}a shows that time distribution ($g_2(\tau,\chi)$) of the bound state with mass-defect exhibits final time-size behaviour. In other words, the final temporal size of the system leads to the observable shift from the mass-shell. To define the temporal size of the system, we take positions of the first zero of the main peak. The following relation defines the positions of the zero: 
\begin{equation}
	(2E-M)\tau_0=\pi.
\label{time-size}\end{equation}
The averge value of the $\tau_0$ is about $1.4\,MeV^{-1}$, what corresponds $276\,Fm$ or $9.2\cdot~10^{-22}$ secounds. 

The Eq.(\ref{time-size}) shows a relation between the off-mass-shell shift and the finite temporal size of the system. Fig.~\ref{D_large_t}b compares the calculations with negative binding energy, zero and small positive binding energy. From the figure, we see that at zero binding energy, the temporal size of the bound state becomes infinite.    

\begin{figure}[!thb]
\begin{center}
\begin{tabular}{cc}
\includegraphics[scale=0.39,clip=true]{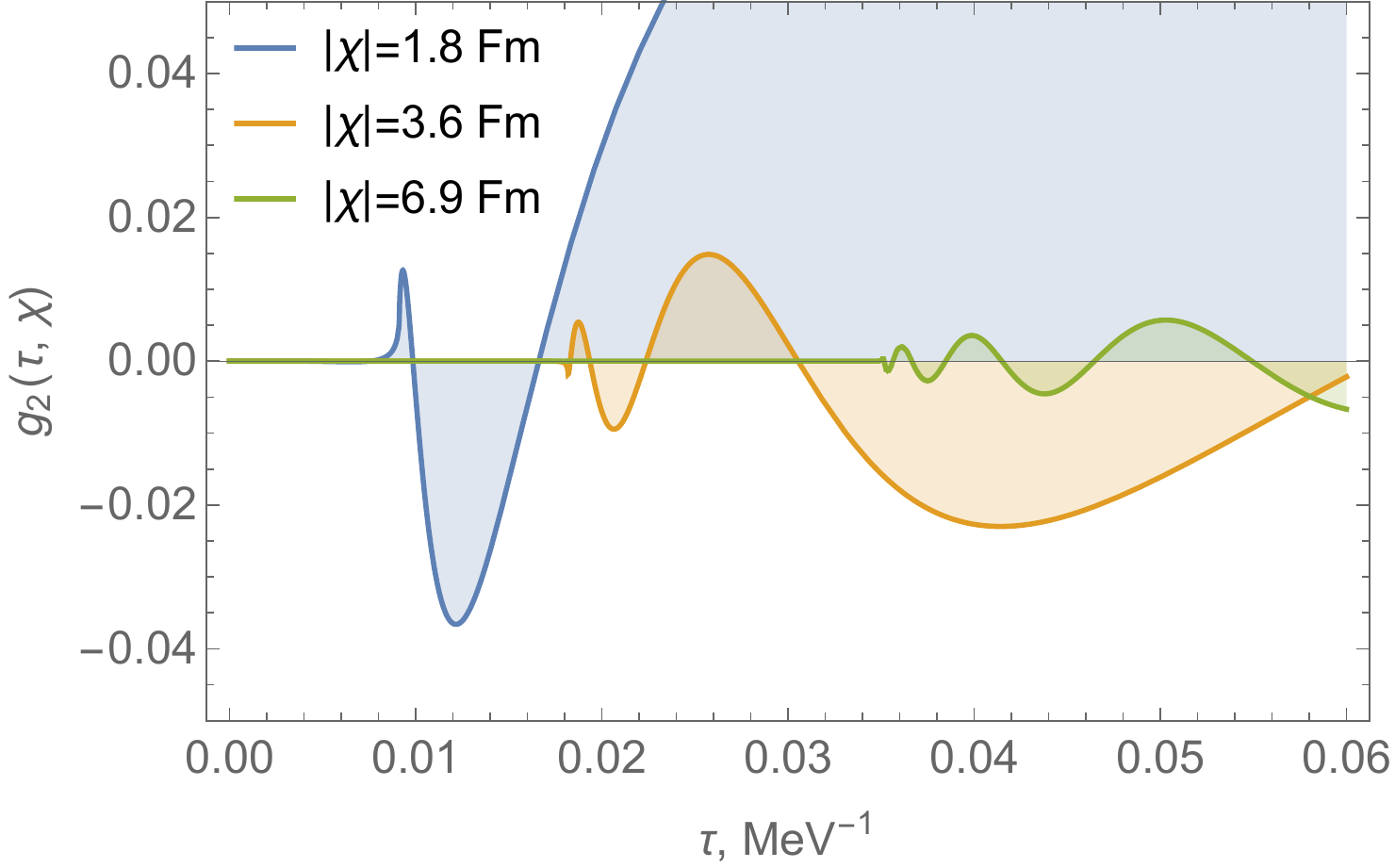} &
\includegraphics[scale=0.41,clip=true]{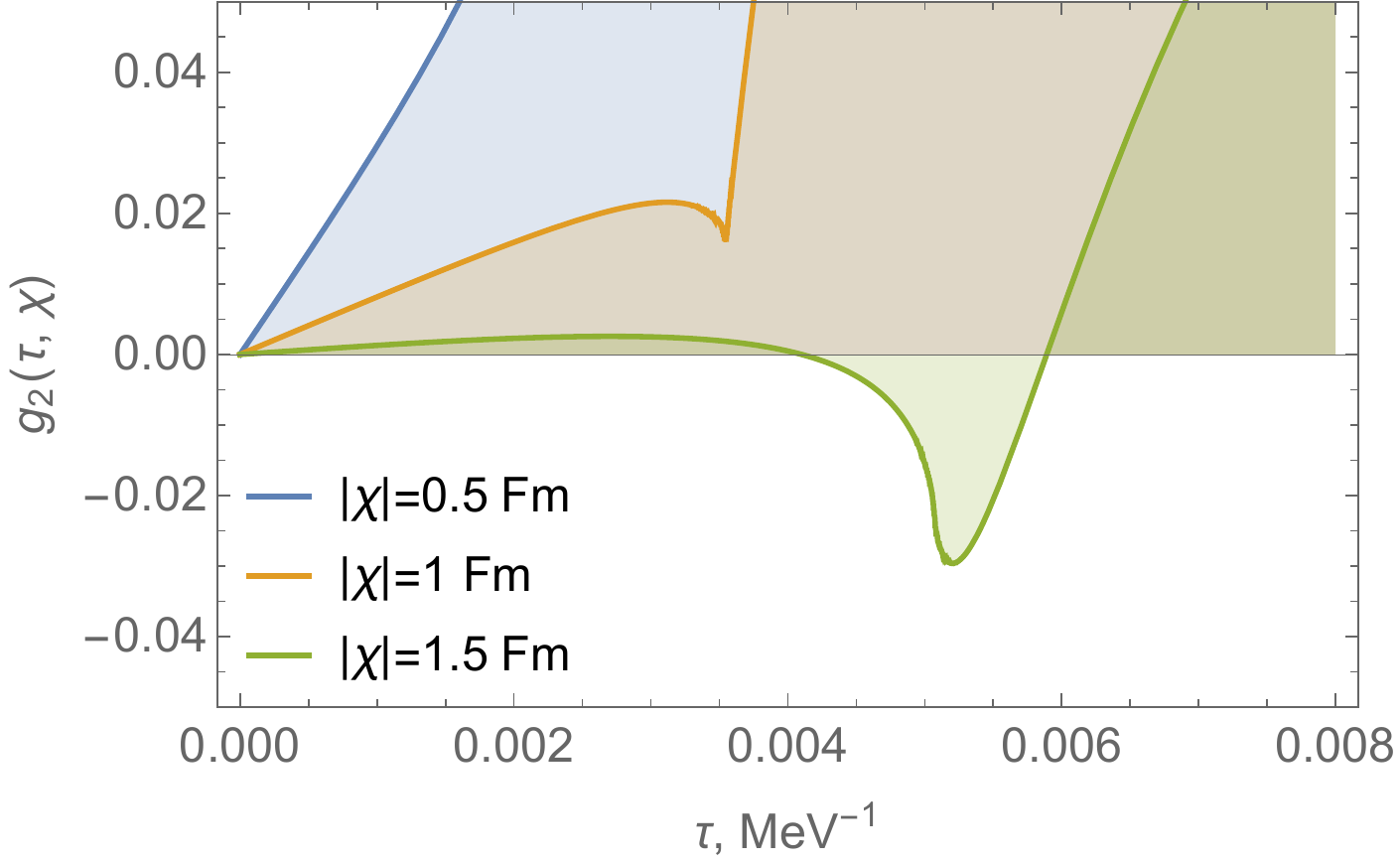}\\
(a) & (b)
\end{tabular}
\end{center}
\caption{(a) The small $\tau$ gap for the three values of space interval - $\chi=1.8 Fm$ (blue line), $\chi=3.6 Fm$ (orange line), $\chi=6.9 Fm$ (green line). The size of the gap corresponds to the light cone condition $\tau^2-\chi^2=0$ \\
(b)The small $\tau$ area with overlaping constituents for the three values of space interval - $\chi=0.5 Fm$ (blue line), $\chi=1 Fm$ (orange line), $\chi=1.5 Fm$ (green line).}
\label{D_small_t}
\end{figure}

Another interesting result is presented in Fig.~\ref{D_small_t}(a). This shows a gap at the small $\tau$ on moderate space distances, where the constituents do not overlap. It means that for the bound non-overlapping particles, the 4-distances with $\tau < \chi$ are forbidden. In the case of small space intervals, where constituents overlap, the gap is blurring (see Fig.~\ref{D_small_t}(b)). 
  
 {\bf EMC-effect} 
 One of the essential questions is - can the finite temporal size of the bound state lead to any observable effects? Here we will consider one - the EMC-effect. 
  
  The EMC-effect discovered in the experiments of the EMC collaboration~\cite{EMC} shows that the short-range structure of a bound nucleon has observable differences from the free nucleon structure. This discovery contradicts the traditional picture of bound states. The scales of nuclear and intranuclear forces differ so much that the former cannot affect the latter.
   
 This contradiction can be solved if one considers the 4D structure of bound states. 
  
 Indeed, since the temporal distribution of a bound nucleon is not uniform and limited, the time translation invariance for an individual constituent is broken\footnote{This circumstance, by the way, explains why quasi-potential approaches were failed to explain the EMC-effect.}. While for a free nucleon, it is conserved. Since the time-invariance is conserved for the whole bound state, the corresponding observables have physical meaning after integration over the bound state space-time.  
 
 The covariant treatment of the deep inelastic amplitude and integration with respect to the zero-component of the relative momentum $k_0$ give the following relation between the bound and free nucleon structure functions: 
 \begin{equation}
 	F_2^{\tilde N}(x)=\int \frac{d^3{\bf k}}{(2\pi)^3} \left[\frac{E-k_3}{E}F^N_2(x_N)+\frac{2E-M}{E}x_N\frac{dF^N_2(x_N)}{dx_N}\right]\Phi^2({\bf k}),
 \label{DIS_SF}\end{equation}
 where $x=Q^2/(2M q_0)$ and $x_N=Q^2/(Eq_0-k_3q_3)$ are Bjorken scaling variable for the deuteron and nucleon respectively. The $Q^2=-q^2=q_3-q_0$ virtual photon space-like momentum.  
 The first term in Eq.(\ref{DIS_SF}) is the contribution of the Fermi-motion. The second term gives finite temporal size effects. 
 
\begin{figure}[!thb]
\begin{center}
\begin{tabular}{cc}
\includegraphics[scale=0.39,clip=true]{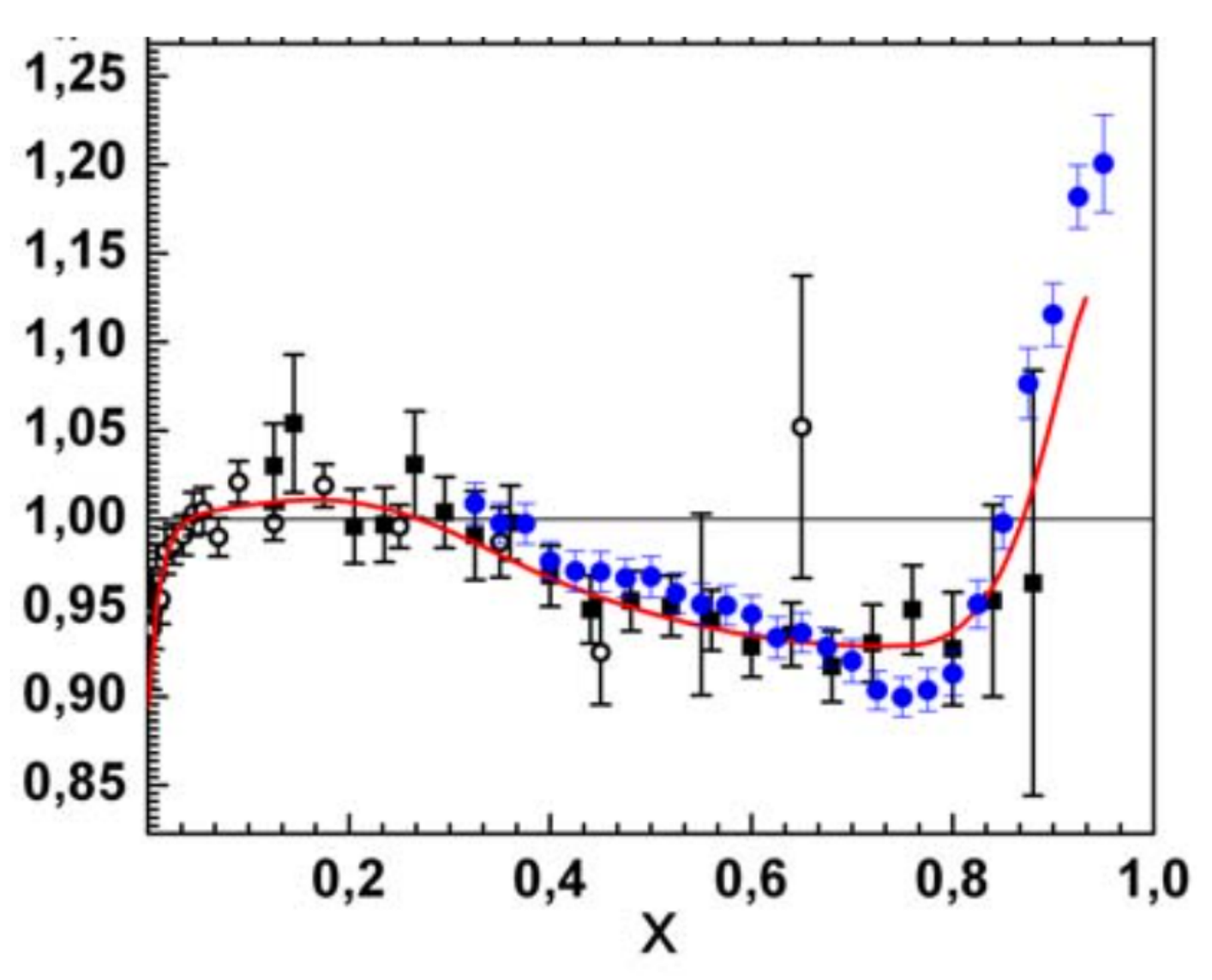} &
\includegraphics[scale=0.291,clip=true]{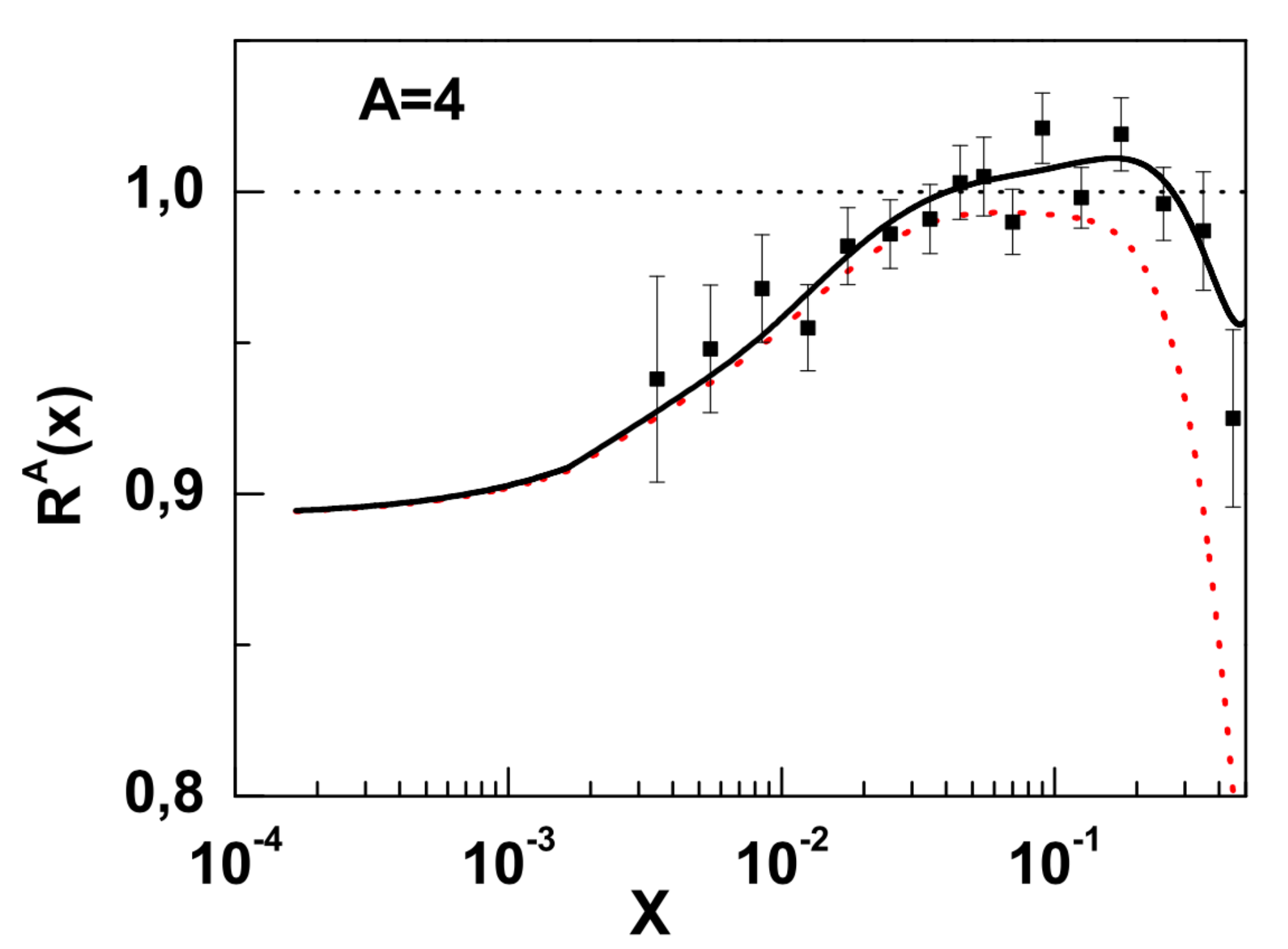}\\
(a) & (b)
\end{tabular}
\end{center}
\caption{(a) Ratio of the 4He and D structure functions.The experimental values are shown by the full squares~\cite{EMC_He1}, the light circles~\cite{EMC_smallx} and full circles~\cite{EMC_He2}. \\
(b)The ratio of the nuclear to deuteron structure functions at small Bjorken x for $^4$He.The experimental values are shown by the solid squares~\cite{EMC_smallx}}
\label{EMC_effect}
\end{figure}

 The previous numerical calculations presented in Fig.~(\ref{EMC_effect}) show good agreement with data in the whole region of $x$ values. 
 
 \section{Conclusion}
 The main question discussed in the presented article is the nature of the vacuum and matter restructuring in spaces with boundaries. We have considered several examples of systems with boundaries to clarify this question. The examples we have considered illustrate three different causes of vacuum and matter restructuring. 
 
 The strongly interacting fermion system demonstrates that the boundary effects restore the chiral symmetry in a chirally broken phase. Due to the boundary effects, the critical temperature decreases, and the phase transition becomes first order. The chiral symmetry is restored at a sufficiently small distance between the plates, even at zero temperature.  
 
 The compact QED with Casimir boundaries demonstrates a purely geometrical effect. The small interplate distance constrains the monopole density, leading to the early deconfinement phase transition. 
 
 The non-Abelian SU(2) gauge field theory demonstrates vacuum restructuring due to the non-perturbative dynamics of gluon fields. It shifts the scale defined by glueball mass to significantly smaller values. It also exhibits early deconfinement phase transition, which might have similar nature as in compact QED. 
 
 The two-fermion bound state exhibits mass scale shift due to the space-like temporal boundaries, where the boundaries have a dynamical nature. It is a purely geometrical effect. The limited temporal size of the bound state leads to the time translation invariance breaking for an individual constituent. It causes the mass-scale shift, or in other words, off-mass-shell effects. One of the observable effects of the finite temporal size of bound states of nucleons is the EMC-effect.  

\section*{Acknowledgments}
I thank M.~Chernodub, and V.~Goy, for fruitful discussions and collaboration on related topics. This work has been supported by the Ministry of Science and Higher Education of Russia (Project No. 0657-2020-0015)


\begin{thebibliography}{99}
\bibitem{Casimir:1948dh} 
  H.~B.~G.~Casimir, {Indag.\ Math.\  {\bf 10}, 261 (1948) [Kon.\ Ned.\ Akad.\ Wetensch.\ Proc.\  {\bf 51}, 793 (1948)]}.

\bibitem{Chernodub:2017mhi} 
  M.~N.~Chernodub, V.~A.~Goy and A.~V.~Molochkov,
  {Phys.\ Rev.\ D {\bf 95}, no. 7, 074511 (2017)}.
  
\bibitem{Chernodub:2017gwe} 
  M.~N.~Chernodub, V.~A.~Goy and A.~V.~Molochkov,
 {Phys.\ Rev.\ D {\bf 96}, no. 9, 094507 (2017)}.
  
\bibitem{Chernodub:2018pmt} 
  M.~N.~Chernodub, V.~A.~Goy, A.~V.~Molochkov and H.~H.~Nguyen,
   {Phys.\ Rev.\ Lett.\  {\bf 121}, no. 19, 191601 (2018)}.

 \bibitem{Flachi:2013bc} 
  A.~Flachi,
 {Phys.\ Rev.\ Lett.\  {\bf 110}, no. 6, 060401 (2013)}.

\bibitem{Flachi:2012pf} 
  A.~Flachi, {Phys.\ Rev.\ D {\bf 86}, 104047 (2012)}.

\bibitem{Tiburzi:2013vza} 
  B.~C.~Tiburzi,
 {Phys.\ Rev.\ D {\bf 88}, 034027 (2013)}
 
 \bibitem{Chernodub:2016kxh} 
  M.~N.~Chernodub and S.~Gongyo,
{Phys.\ Rev.\ D {\bf 95}, no. 9, 096006 (2017)}.


\bibitem{ref:Bogdag}
M. Bordag, G. L. Klimchitskaya, U. Mohideen, and V. M. Mostepanenko, 
Advances in the Casimir Effect (Oxford University Press, New York, 2009).

\bibitem{ref:Milton}
K. A. Milton,
The Casimir Effect: Physical Manifestations of Zero-Point Energy
(World Scientific Publishing, Singapore, 2001).

\bibitem{Lamoreaux:1996wh} 
  S.~K.~Lamoreaux,
 {Phys.\ Rev.\ Lett.\  {\bf 78}, 5 (1997)}; Erratum: [{Phys.\ Rev.\ Lett.\  {\bf 81}, 5475 (1998)}].

\bibitem{Mohideen:1998iz}
  U.~Mohideen and A.~Roy,
  {Phys.\ Rev.\ Lett.\  {\bf 81}, 4549 (1998)}
\bibitem{Boyer}
T. H. Boyer, Phys. Rev. 174, 1764 (1968)


\bibitem{Scharnhorst:1990sr} 
  K.~Scharnhorst, {Phys.\ Lett.\ B {\bf 236}, 354 (1990)}.
\bibitem{Scharnhorst:1993} G.~Barton and K.~Scharnhorst 1993 J. Phys. A: Math. Gen. 26 2037
\bibitem{Chernodub:2019nct}
M.~N.~Chernodub, V.~A.~Goy and A.~V.~Molochkov, PoS \textbf{Confinement2018}, 006 (2019)   
\bibitem{ref:BKT}
V.~L.~Berezinskii,
{Sov. Phys. JETP {\bf 32}, 493 (1970)}; 
{\it $\dots$ II. Quantum Systems},
{Sov. Phys. JETP {\bf 34}, 610 (1971)};
  J.~M.~Kosterlitz and D.~J.~Thouless,
 {J.\ Phys.\ C {\bf 6}, 1181 (1973)}.

\bibitem{ref:binding:cU1} 
  M.~N.~Chernodub, E.~M.~Ilgenfritz and A.~Schiller,
  {Phys.\ Rev.\ D {\bf 64}, 054507 (2001)}.

\bibitem{Teper:1998te} 
  M.~J.~Teper,
 {Phys.\ Rev.\ D {\bf 59}, 014512 (1998)}
  A.~Athenodorou and M.~Teper, {JHEP {\bf 1702}, 015 (2017)}
  
\bibitem{Karabali:2018ael} 
  D.~Karabali and V.~P.~Nair, {Phys.\ Rev.\ D {\bf 98}, no. 10, 105009 (2018)}.

\bibitem{BS} E.E. Salpeter, H.A. Bethe, Phys. Rev. 84, 1232 (1951).

\bibitem{EMC}
J.J. Aubert et al., Phys. Lett. B123, 275 (1983).

\bibitem{EMC_He1} 
J. Gomez et al., Phys. Rev. D49, 4348 (1994).

\bibitem{EMC_smallx}
NMC, P. Amaudruz, et al., Nucl.Phys. B441, 3 (1995).

\bibitem{EMC_He2}
J. Seely et al., Phys. Rev. Lett. 103, 202301 (2009).

\end{thebibliography}
\end{document}